\documentclass[twocolumn,citeautoscript,preprintnumbers,amsmath,amssymb,showpacs,prb]{revtex4}
\usepackage{graphicx}
\usepackage{dcolumn}
\usepackage{bm}
\usepackage{times}

\newcommand{\ca}{CaFe$_{2}$As$_{2}$}

\newcommand{\pt}{$p-$T}

\begin{document}
\title{Lattice collapse and quenching of magnetism in CaFe$_{2}$As$_{2}$ under pressure: A single crystal neutron and x-ray diffraction investigation}

\author{A.~I.~Goldman$^{1,2}$, A.~Kreyssig$^{1,2}$, K.~Proke\v{s}$^{3}$,
D.~K.~Pratt$^{1,2}$, D.~N.~Argyriou$^{3}$, J.~W.~Lynn$^{4}$,
S.~Nandi$^{1,2}$, S.~A.~J.~Kimber $^{3}$, Y.~Chen$^{4,5}$,
Y.~B.~Lee$^{1,2}$, G.~Samolyuk$^{1,2}$, J.~B.~Le\~{a}o$^{4}$,
S.~J.~Poulton$^{4,5}$, S.~L.~Bud'ko$^{1,2}$, N.~Ni$^{1,2}$,
P.~C.~Canfield$^{1,2}$ B.~N.~Harmon$^{1,2}$ and R.~J.~McQueeney$^{1,2}$} \affiliation{\\
$^1$ Ames Laboratory, US DOE, Iowa State University, Ames, IA 50011,
USA \\$^2$Department of Physics and Astronomy, Iowa State
University, Ames, IA 50011, USA\\$^3$Helmholtz-Zentrum Berlin
f\"{u}r Materialien und Energie, Glienicker Str. 100, 14109 Berlin,
Germany\\$^4$NIST Center for Neutron Research, National Institute of
Standards and Technology, Gaithersburg, MD 20899,
USA\\$^5$Department of Materials Science and Engineering, University
of Maryland, College Park, MD 20742, USA}

\date{\today}
\begin{abstract}
Single crystal neutron and high-energy x-ray diffraction have
identified the phase lines corresponding to transitions between the
ambient-pressure tetragonal (T), the antiferromagnetic orthorhombic
(O) and the non-magnetic collapsed tetragonal (cT) phases of
CaFe$_{2}$As$_{2}$. We find no evidence of additional structures for
pressures up to 2.5~GPa (at 300~K). Both the T-cT and O-cT
transitions exhibit significant hysteresis effects and we
demonstrate that coexistence of the O and cT phases can occur if a
non-hydrostatic component of pressure is present.  Measurements of
the magnetic diffraction peaks show no change in the magnetic
structure or ordered moment as a function of pressure in the O phase
and we find no evidence of magnetic ordering in the cT phase.
Band structure calculations show that the transition results in a
strong decrease of the iron 3d density of states at the Fermi
energy, consistent with a loss of the magnetic moment.
\end{abstract}

\maketitle

The discovery\cite{torikachvili,park} of pressure-induced
superconductivity in CaFe$_{2}$As$_{2}$ has opened an exciting new
avenue for investigations of the relationship between magnetism,
superconductivity, and lattice instabilities in the iron arsenide
family of superconductors. Features found in the compositional phase
diagrams of the iron arsenides\cite{norman}, such as a
superconducting region at low temperature and finite doping
concentrations, are mirrored in the pressure-temperature phase
diagrams. Superconductivity appears at either a critical doping, or
above some critical pressure in the \textit{A}Fe$_{2}$As$_{2}$
(\textit{A}$=$Ba,~Sr,~Ca) or '122' family of compounds, raising
questions regarding the role of both electronic doping and pressure,
especially in light of the recent observation of pressure induced
superconductivity in the related compound, LaFeAsO\cite{okada}. Does
doping simply add charge carriers, or are changes in the chemical
pressure, upon doping, important as well? What subtle, or striking,
modifications in structure or magnetism occur with doping or
pressure, and how are they related to superconductivity?

Similar to other members of the \textit{A}Fe$_{2}$As$_{2}$
(\textit{A}$=$Ba, Sr) family\cite{huang,jesche,zhao,su}, at ambient
pressure CaFe$_{2}$As$_{2}$ undergoes a transition from a
non-magnetically ordered tetragonal (T) phase (a~=~3.879(3)~{\AA},
c~=~11.740(3)~{\AA}) to an antiferromagnetic (AF) orthorhombic (O)
phase (a~=~5.5312(2)~{\AA}, b~=~5.4576(2)~{\AA},
c~=~11.683(1)~{\AA}) below approximately 170~K.\cite{ni,goldman} In
the O phase, Fe moments order in the so called AF2
structure\cite{yildirim2} with moments directed along the
\emph{a}-axis of the orthorhombic structure\cite{goldman}. Neutron
powder diffraction measurements\cite{kreyssig} of CaFe$_{2}$As$_{2}$
under hydrostatic pressure found that for \textit{p}$>$0.35~GPa (at
\textit{T}$=$50~K), the antiferromagnetic O phase transforms to a
new, non-magnetically ordered, collapsed tetragonal (cT) structure
(a~=~3.9792(1)~{\AA}, c~=~10.6379(6)~{\AA}) with a dramatic decrease
in both the unit cell volume (5\%) and the \textit{c/a} ratio
(11\%). The transition to the cT phase occurs in close proximity to
the pressure at which superconductivity is first
observed\cite{torikachvili}. Total energy calculations based on this
cT structure concluded that the Fe moment is quenched, consistent
with the absence of magnetic neutron diffraction
peaks\cite{kreyssig}.

Since the report of a cT phase in CaFe$_{2}$As$_{2}$ under pressure,
a significant effort has been devoted to understanding the
relationship between the cT structure, features observed in
resistivity and susceptibility measurements\cite{torikachvili,lee},
and the results of local probe measurements such as
$\mu$SR\cite{goko}. Although the neutron powder diffraction
measurements demonstrated the loss of magnetic order in the cT
phase, subsequent theoretical work has proposed that the Fe moment
in the cT phase is not quenched and orders in an alternative N\`eel
state (the so-called AF1 structure)\cite{yildirim}. It has also been
suggested, from $\mu$SR measurements, that a partial volume fraction
of static magnetic order coexists with superconductivity over an
intermediate pressure range\cite{goko}. Most recently, the existence
of a new structural phase above 0.75~GPa in CaFe$_{2}$As$_{2}$
(Phase \textrm{I\!I\!I} in Ref.~~\onlinecite{lee}) has been proposed
based on resistivity measurements.  It is, therefore, important to
clearly identify the chemical and magnetic structures of \ca, and
the phase lines that separate them as a function of temperature and
pressure, a task best accomplished by neutron and x-ray diffraction
measurements.

In this paper, we report on neutron and x-ray single crystal
diffraction studies of the pressure-temperature (\pt) magnetic and
structural phase diagram of \ca. We clearly  identify the phase
boundaries of the T, O and cT phases in the $p-$T phase diagram and
find significant hysteresis associated with transitions to, and
from, the cT phase.  We find no evidence of additional structural
phases\cite{lee} for pressures up to 2.5GPa.  We also demonstrate
that the AF2 ordering is associated only with the O phase, and
vanishes upon entering the cT phase in agreement with previous
powder diffraction measurements\cite{kreyssig}.  The apparent
coexistence between magnetic order and superconductivity under
pressure\cite{goko} most likely arises from coexistence between the
O and cT phases under non-hydrostatic measurement conditions.
Furthermore, measurements at reciprocal lattice points associated
with one of the three-dimensional realizations of the proposed AF1
structure\cite{yildirim} does not reveal any evidence of magnetic
order in the cT phase. Finally, we show from band structure
calculations that the collapse of the lattice leads to a strong
reduction of the Fe 3d density of states at the Fermi energy,
consistent with the loss of magnetism in the cT phase.

\section {Experimental Details}
The single crystals of CaFe$_{2}$As$_{2}$ used for the diffraction
measurements were grown either using a Sn flux as described
previously\cite{ni}or from an FeAs flux.  The FeAs powder used for
the self-flux growth was synthesized by reacting Fe and As powders
after they were mixed and pressed into pellets. The pellets were
sealed inside a quartz tube under one-third of an atmosphere of Ar
gas, slowly heated to 500$^\circ$~C, held at that temperature for
ten hours, and then slowly heated to 900$^\circ$~C and held at
900$^\circ$~C for an additional ten hours.  Single crystals of
CaFe$_{2}$As$_{2}$ were grown from this self flux using conventional
high-temperature solution growth techniques. Small Ca chunks and the
FeAs powder were mixed together in a 1:4 ratio. The mixture was
placed into an alumina crucible, together with a second catch
crucible containing quartz wool, and sealed in a quartz tube under
one-third of an atmosphere of Ar gas. The sealed quartz tube was
heated to 1180$^\circ$~C for 2 hours, cooled to 1020$^\circ$~C over
4 hours, and then slowly cooled to 970$^\circ$~C over 27 hours where
the FeAs was decanted from the plate like single crystals. The
as-grown crystals were annealed at 500$^\circ$~C for 24 hours.

Neutron diffraction data were taken on the BT-7 spectrometer at the
NIST Center for Neutron Research and on the E4 diffractometer at the
Helmholtz-Zentrum Berlin f\"{u}r Materialien und Energie.
High-energy x-ray diffraction data were acquired using station
6-ID-D in the MUCAT Sector at the Advanced Photon Source.

Since much of the description below involves diffraction
measurements in both the tetragonal and orthorhombic phases of \ca,
it is useful to describe the indexing system employed in our
discussions. For the orthorhombic structure we employ indices
(\textit{HKL})$_{\textup{O}}$ for the reflections based on the
relations, \emph{H~=~h~+~k}, \emph{K~=~h~-~k} and \emph{L~=~l},
where (\textit{hkl})$_{\textup{T}}$ are the corresponding Miller
indices for the tetragonal phase.  For example, the
(220)$_{\textup{T}}$ tetragonal peak becomes the orthorhombic
(400)$_{\textup{O}}$ below the structural transition. The
(\textit{hhl})$_{\textup{T}}$ reciprocal lattice plane in the
tetragonal structure becomes the (\textit{H0L})$_{\textup{O}}$ plane
in the orthorhombic phase. The indexing notation for the collapsed
tetragonal phase is identical to that for the tetragonal structure.
When referring to diffraction peaks, we will use subscripts T, O and
cT to denote whether the peaks are indexed in the tetragonal,
orthorhombic or collapsed tetragonal phase.

\emph{Neutron diffraction measurements on BT-7}: Three sets of
measurements were performed on the BT-7 diffractometer.  The first
data set focused on measurements of phase boundaries separating the
T, O and cT phase. These data were taken in double-axis mode, using
a wavelength of $\lambda=$2.36~\AA~and two pyrolytic graphite
filters to reduce higher harmonic content of the beam. A 10 mg
single crystal (3$\times$3$\times$0.2~mm$^3$), wrapped in Al-foil
was secured to a flat plate within the Al-alloy He-gas pressure cell
and cooled using a closed-cycle refrigerator. The sample was
oriented so that the (\textit{hhl})$_{\textup{T}}$ reciprocal
lattice plane was coincident with the scattering plane of the
diffractometer. The second set of data focused on measurements of
the magnetic scattering in the orthorhombic phase.  To maximize the
intensity of the magnetic scattering relative to the substantial
background from the pressure cell, these data were taken on a
composite of eight single crystals, attached to a support plate
using Fomblin oil, with the diffractometer operated in triple-axis
mode using a PG(002) analyzer. The single crystals (combined mass of
approximately 60 mg) were co-aligned so that their common
(\textit{hhl})$_{\textup{T}}$ plane was aligned in the scattering
plane. The mosaic of the composite sample with respect to the
[\textit{hh0}]$_{\textup{T}}$ and [\textit{00l}]$_{\textup{T}}$
directions was approximately 1 deg full-width-at-half-maximum
(FWHM). The third data set focused on a survey of magnetic
scattering vectors in the (\textit{h0l})$_{\textup{cT}}$ diffraction
plane of the cT phase associated with the proposed AF1 structure
using the same diffractometer conditions as for data set two. For
these measurements, a 70 mg single crystal, grown in an FeAs flux,
was wrapped in Al-foil and secured to a flat plate within the
Al-alloy He-gas pressure cell. The crystal mosaic of this sample was
measured to be approximately 1 deg (FWHM).

All of these measurements employed an Al-alloy He-gas pressure cell
to ensure hydrostatic pressure conditions.  The cell was connected
to a pressurizing intensifier through a high pressure capillary that
allowed continuous monitoring and adjusting of the pressure. Using
this system, the pressure could be varied at fixed temperatures
(above the He solidification line), or the temperature could be
scanned at nearly constant pressures. A helium reservoir allowed the
pressure to remain relatively constant as temperature was changed.
The finite size of the reservoir, however, results in some change in
pressure over the temperature range measured, on the order of 15\%
for the highest pressures ($\sim$0.6~GPa).

\emph{Neutron diffraction measurements on E4}: Additional neutron
diffraction data were measured up to a pressure of 1 GPa on the E4
double axis diffractometer with $\lambda=$2.44~\AA\ for the
(\textit{hhl)$_{\textup{T}}$} orientation of a 10 mg Sn-flux grown
single crystal. Here we employed a Be-Cu clamp-type pressure cell
and a 1:1 mixture of Fluorinert 77 and 70, for the pressure medium.
The pressure was set \textit{ex-situ} at room temperature and the
cell was inserted into a standard He cryostat. The initial pressure
was measured using a manganin pressure sensor and then monitored
\textit{in-situ} (since pressure decreases with decreasing
temperature) by tracking the lattice constants of NaCl crystals
placed above the sample.

\emph{High-energy x-ray diffraction measurements}: In order to move
beyond the limits of the He-gas and clamped pressure cells we used a
Merrill-Bassett diamond anvil pressure cell with an ethanol/methanol
mixture as the pressure medium. This cell allowed us to collect
diffraction data at 300K up to 2.5 GPa. Here we employed 100~keV
x-rays to ensure full penetration of single crystal (Sn-flux grown)
samples in the cell and recorded diffraction data over layers of
reciprocal space using a 2D detector. At all pressures investigated
the medium remained fluid.

\section {Results}

\subsection{Determination of the pressure-temperature phase diagram}
\begin{figure*}
\begin{center}
\includegraphics[clip, width=1.0\textwidth]{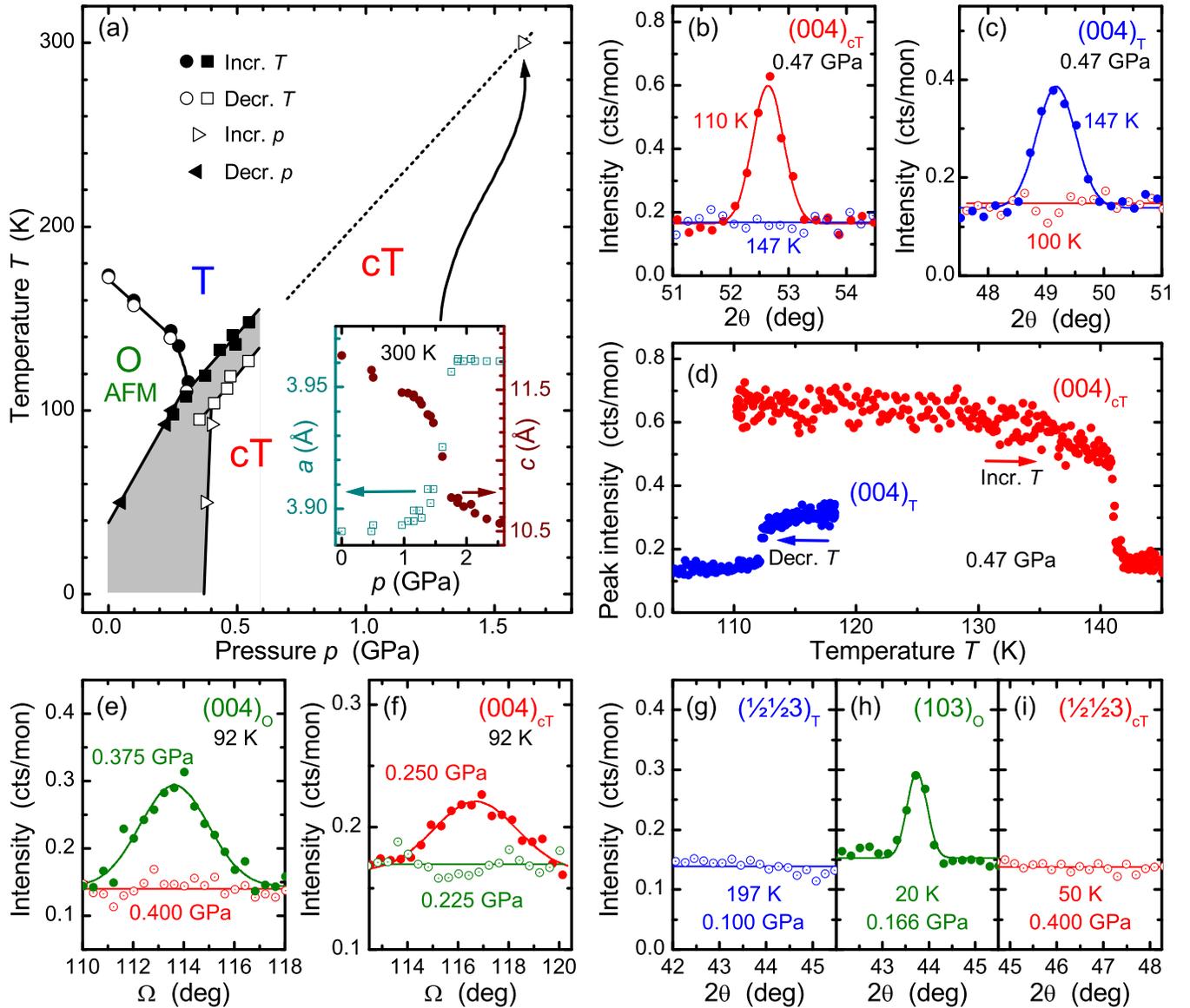}\\
\caption{(color online) (a) Pressure-temperature phase diagram of
CaFe$_{2}$As$_{2}$ under hydrostatic pressure determined from
neutron and high-energy x-ray diffraction measurements. Filled and
open circles (squares) denote phase boundaries determined upon
heating and cooling at a set pressure for the O-T (cT-T) phase
transition, respectively. Filled and open triangles denote phase
boundaries determined upon increasing and decreasing pressure at a
fixed temperature, respectively. The shaded area denotes the
hysteretic region.  The inset of Fig. 1(a) shows the change in
lattice constants at the T-cT transition at 300 K as measured by
high-energy x-ray diffraction. In Figs.~\ref{fig:fig1}(b)-(i), the
color codes denote measurements of the T phase (blue), O phase
(green) or cT phase (red) diffraction peak positions. (b)
$\Omega$-2$\theta$ scans of the (004)$_{\textup{cT}}$ peak at 0.47
GPa on increasing temperature. Here $\Omega$ denotes the sample
angle while 2$\theta$ is the scattering angle. (c)
$\Omega$-2$\theta$ scans of the (004)$_{\textup{T}}$ peak at 0.47
GPa on decreasing temperature. (d) Temperature dependence of the
peak intensities of the (004)$_{\textup{T}}$ as temperature is
decreased through the T-cT transition at \textit{p}=0.47 GPa, and
the (004)$_{\textup{cT}}$ as temperature is increased through the
T-cT transition at \textit{p}=0.47 GPa. (e) $\Omega$ scans through
the (004)$_{\textup{O}}$ peak at 92 K as the pressure is increased
from 0.375 to 0.400 GPa. (f) $\Omega$ scans through the
(004)$_{\textup{cT}}$ peak at 92 K as the pressure is decreased from
0.250 to 0.225 GPa. (g) $\Omega$-2$\theta$ scans through the
expected position of the
($\frac{1}{2}$$\frac{1}{2}$3)$_{\textup{T}}$ magnetic peak in the
tetragonal phase. (h) $\Omega$-2$\theta$ scans through the observed
position of the (103)$_{\textup{O}}$ magnetic peak in the
orthorhombic phase. (i) $\Omega$-2$\theta$  scans through the
expected position of the
($\frac{1}{2}$$\frac{1}{2}$3)$_{\textup{cT}}$ magnetic peak in the
collapsed tetragonal phase.} \label{fig:fig1}
\end{center}
\end{figure*}

We first describe how the temperature/pressure phase lines for
CaFe$_{2}$As$_{2}$, shown in Fig.~\ref{fig:fig1}(a), were derived
from our neutron and high-energy x-ray diffraction measurements. The
T-O phase boundary was determined by monitoring the
(220)$_{\textup{T}}$ or (112)$_{\textup{T}}$ diffraction peaks upon
heating (solid circles) or cooling (open circles) at specified
pressures. At the T-O transition, the orthorhombic distortion and
associated twinning splits both the (220)$_{\textup{T}}$ and
(112)$_{\textup{T}}$ peaks, providing a clear signature of the
transition. As indicated in Fig.~\ref{fig:fig1}(a), there is little
hysteresis in the T-O transition ($\sim$2-3~K), consistent with our
previous measurements at ambient
pressure\cite{torikachvili,ni,goldman}.

In a similar fashion, the T-cT phase boundary in
Fig.~\ref{fig:fig1}(a) was mapped by monitoring the intensity of the
(004)$_{\textup{cT}}$ peak while heating the sample from low
temperature (solid squares), or the intensity of the
(004)$_{\textup{T}}$ diffraction peak while cooling the sample from
higher temperature (open squares). The sizeable difference
($\sim$9\%) in the \textit{c}-lattice constants for these two
structures is evident from the scattering angles for the two peaks
in Figs.~\ref{fig:fig1}(b) and (c). Figure~\ref{fig:fig1}(d) shows
that, at a nominal pressure of 0.47~GPa, the cT phase transforms
sharply to the T phase at 141~K (on heating) with no coexistence
beyond the 1~K wide transition region. On decreasing temperature,
the transition from the T to the cT phase is also very sharp but
occurs at 112~K, nearly 30~K below the transition on heating,
demonstrating the strong hysteresis (similar in magnitude noted in
Refs.~~\onlinecite{torikachvili} and~~\onlinecite{lee}) in the T-cT
transition. We also point out that the strong volume changes
associated with the T-cT transition irresversibly increases the
sample mosaic. Indeed, the difference between the measured peak
intensities for the (004) diffraction peaks in
Fig.~\ref{fig:fig1}(d) arises from the doubling of the sample mosaic
during this transition.  We note, however, that all transitions in
temperature and pressure were reproducible, even after several
pressure/temperature cycles.

The phase boundary between the O and the cT phases was determined by
increasing/decreasing pressure at a fixed temperature while
monitoring the (004)$_{\textup{O}}$ diffraction peak (increasing
pressure, solid triangles) and the (004)$_{\textup{cT}}$ peak
(decreasing pressure, open triangles). For example, at 92~K, on
increasing pressure, a sharp transition from the O to cT phase, with
a width smaller than our step size of 0.025 GPa, was found between
0.375 and 0.400~GPa (Fig.~\ref{fig:fig1}(e)). Upon decreasing
pressure at 92~K, however, the transition to the O phase occurred at
0.225~GPa (Fig.~\ref{fig:fig1}(f)), demonstrating a striking
pressure hysteresis in the O-cT transition as well. Nevertheless,
the transitions themselves were sharp with no phase coexistence in
evidence. We note that above 0.4~GPa, the cT phase persists down to
the lowest temperature (4~K) and highest pressure (0.6~GPa) attained
for the BT-7 measurements.

Based upon resistivity measurements performed in a clamp-type Be-Cu
cell using a silicon fluid as a pressure medium, Lee \textit{et
al.}\cite{lee} have proposed that there is a transition at
approximately 0.75 GPa from the cT phase to another structure of
unknown symmetry (Phase \textrm{I\!I\!I} in Ref.~~\onlinecite{lee}).
In order to move beyond the limits of the He-gas pressure cell to
investigate the possibility of additional phases at higher pressure,
x-ray diffraction measurements were performed using a diamond-anvil
pressure cell at 300~K on the 6ID-D station in the MUCAT Sector at
the Advanced Photon Source. Diffraction data over a wide range of
reciprocal space in both the (\textit{hhl})$_{\textup{T}}$ and
(\textit{h0l})$_{\textup{T}}$ planes were collected using a
two-dimensional area detector. As described in earlier
work\cite{kreyssig2}, entire reciprocal lattice planes can be imaged
to identify any new reflections that signal the presence of
additional structural phases. The T-cT transition at 300~K is
evidenced by the strong change in lattice constants observed at
1.6~GPa, as shown in the inset of Fig.~\ref{fig:fig1}(a). No
additional diffraction peaks were found above this transition,
confirming the presence of only the cT phase for pressures up to 2.5
GPa.

One of the most striking features of Fig.~\ref{fig:fig1}(a) is the
large hysteresis regime represented by the shading.  Within this
area, the structure and physical properties measured at a particular
pressure and temperature depend strongly upon the path taken to that
point. Taken together, this large range of hysteresis and the
strongly anisotropic response of the structure at the cT phase
boundaries (see discussion in Section IIC) can easily lead to
discrepencies in reports of magnetic order, electronic properties
and superconductivity in CaFe$_{2}$As$_{2}$ under pressure.

\subsection{The magnetic structure of \ca under pressure}
\begin{figure}
\begin{center}
\includegraphics[clip, width=.45\textwidth]{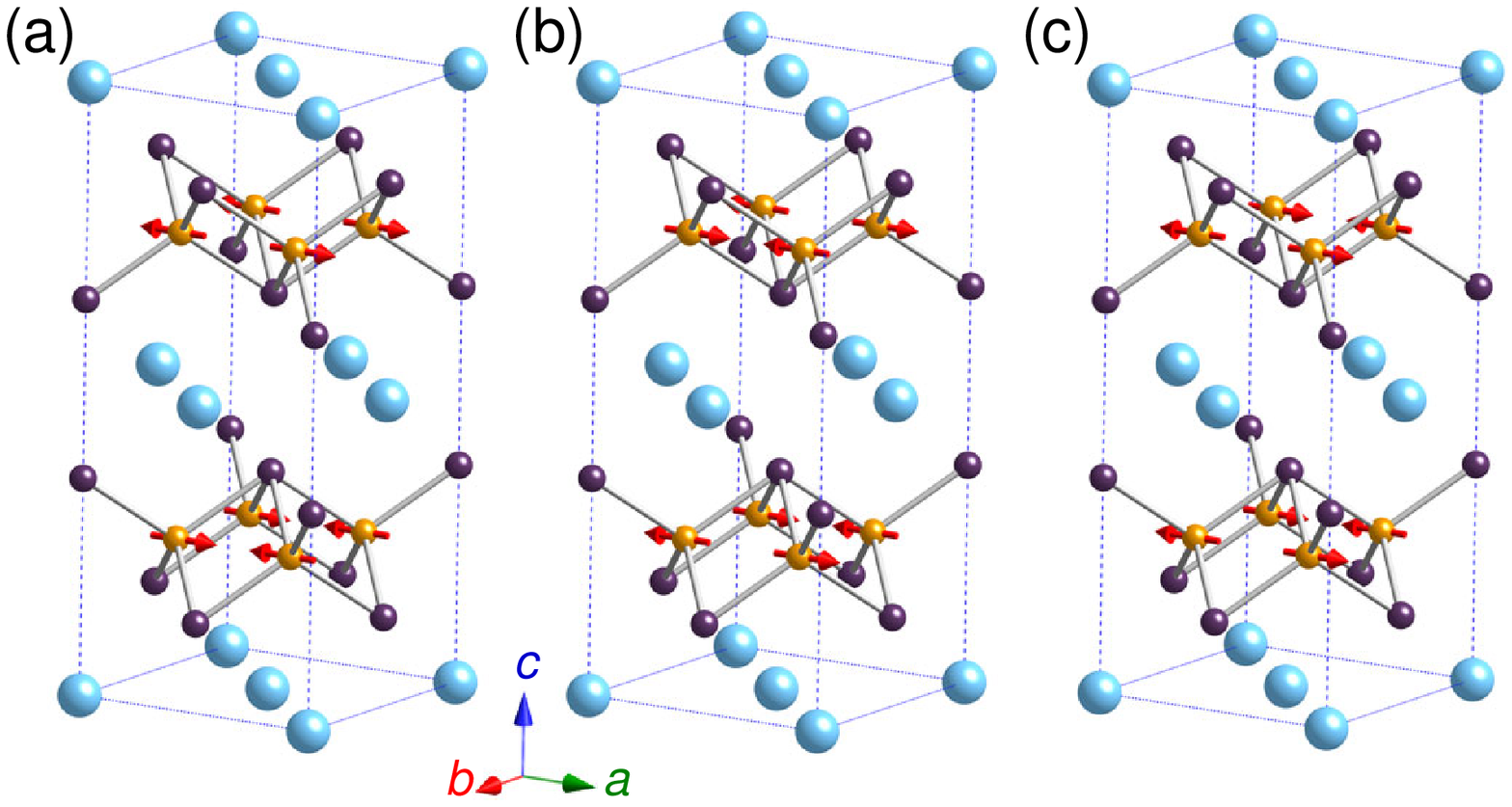}
\caption{(color online) Magnetic structures for \ca shown in the
orthorhombic unit cell to facilitate comparison. (a) The AF2
magnetic structure realized at ambient pressure below
170~K\cite{goldman}.(b) Representation of the AF1 G-type
antiferromagnetic structure with antiferromagnetic ordering along
the c-axis. (c) Representation of the AF1 C-type antiferromagnetic
structure with ferromagnetic ordering along the
c-axis.}\label{fig:fig2}
\end{center}
\end{figure}

One of the most interesting results of the original report of a cT
phase from neutron powder diffraction measurements\cite{kreyssig},
was the disappearance of Fe magnetic ordering.  The
antiferromagnetic AF2 ordering in the orthorhombic phase of \ca is
shown in Fig.~\ref{fig:fig2}(a).  Magnetic diffraction peaks are
found at positions (\textit{H0L})$_{\textup{O}}$ with \emph{H} and
\emph{L} odd, such as (101)$_{\textup{O}}$ and (103)$_{\textup{O}}$.
During the course of our measurements of the single crystal on BT-7
the magnetic peaks were also monitored. The (101)$_{\textup{O}}$ and
(103)$_{\textup{O}}$ magnetic peaks were observed (see, for example,
Fig.~\ref{fig:fig1}(h)) at selected pressures and temperatures
within the O phase, but no AF2 magnetic peaks were found at the
corresponding ($\frac{1}{2}\frac{1}{2}$1) or
($\frac{1}{2}\frac{1}{2}3$) positions (see Figs.~\ref{fig:fig1}(g)
and (i)) in either the T or cT phases, consistent with previous
results\cite{kreyssig}.

In order to more closely track the evolution of the magnetic
structure with pressure, a composite sample of eight single
crystals, as described in the Section I, was mounted in the He-gas
pressure cell and cooled through the T-O transition at ambient
pressure.  At T = 75~K, pressure was increased in 0.05~GPa steps,
through the O-cT transition, up to a maximum pressure of 0.45~GPa.
The temperature was then lowered to 50~K and pressure was released
in 0.05~GPa steps to ambient pressure.  At each pressure step , the
(004)$_{\textup{O}}$ and (004)$_{\textup{cT}}$ nuclear peaks were
measured along with the (101)$_{\textup{O}}$ and
(103)$_{\textup{O}}$ magnetic peaks.

\begin{figure}
\begin{center}
\includegraphics[clip, width=.45\textwidth]{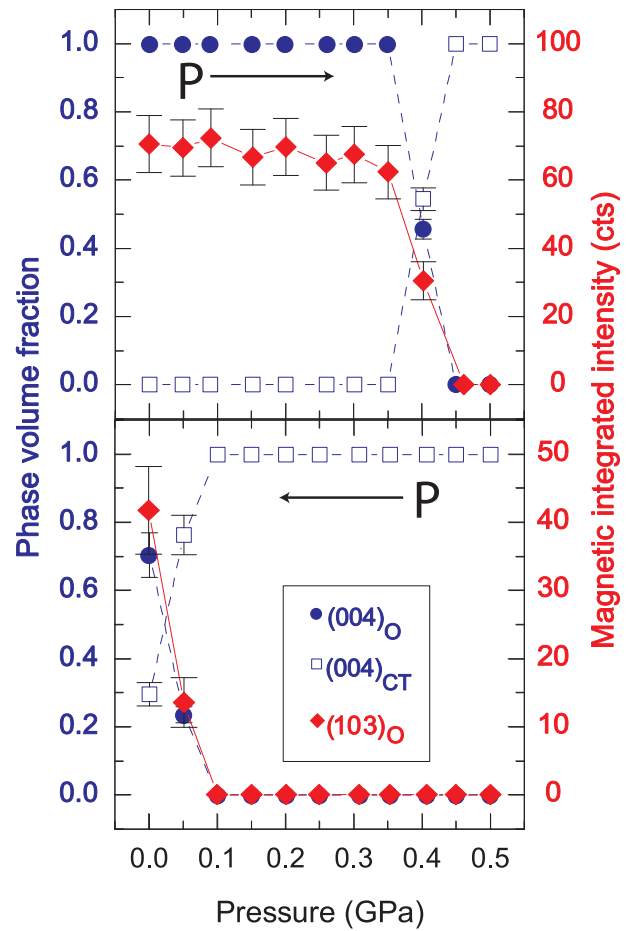}\\
\caption{ (color online) Phase fractions of the O and cT phases as a
function of pressure upon increasing pressure (top panel) at 75~K
and decreasing pressure (bottom panel) at 50~K.  The integrated
intensity of the magnetic (103)$_{\textup{O}}$ reflection remains
constant in the O phase.} \label{fig:fig3}
\end{center}
\end{figure}

Fig.~\ref{fig:fig3} plots the volume fraction of the O and cT phases
as a function of pressure at these two temperatures, determined from
the integrated intensities of the (004)$_{\textup{O}}$ and
(004)$_{\textup{cT}}$ peaks.  The integrated intensity of the
(103)$_{\textup{O}}$ magnetic peak is also plotted at each pressure
value.  Upon increasing pressure at 75~K, the magnetic peak
intensity remains constant until the O-cT phase boundary is reached.
At 0.40~GPa we observe coexistence between the O and cT phases and a
decrease in the magnetic intensity consistent with the decrease in O
phase volume fraction.  For pressures greater than 0.45~GPa, the
sample has completely transformed to the cT phase and there is no
evidence of static antiferromagnetic order associated with the AF2
structure at either the (103)$_{\textup{O}}$ or
($\frac{1}{2}\frac{1}{2}3$)$_{\textup{cT}}$ reciprocal lattice
positions. As pressure is decreased at 50~K, only the cT phase is
present until the cT-O transition boundary at 0.075~GPa, consistent
with the results shown in Fig.~\ref{fig:fig1}(a).  Upon the
appearance of the O phase, the AF2 structure is once again
recovered.

One interesting difference between the data in Fig.~\ref{fig:fig3}
and Fig.~\ref{fig:fig1} is the finite range of coexistence between
the O and cT phases.  In particular, we see that for the composite
sample, the cT phase is evident down to ambient pressure whereas for
the single crystal sample, the transition between the cT and O
phases was sharp, exhibiting little in the way of phase coexistence.
We attribute this to the fact that the composite sample was set on a
sample holder using Fomblin oil, which solidifies well above the
temperatures of these measurements. As described below, dramatic
changes in the unit cell dimensions, as the sample transforms into
(or out of) the cT phase, can introduce significant strain for
constrained samples that "smears" the transition.  Nevertheless,
these data show that: (1) the AF2 magnetic structure is associated
with the O phase and is absent in the cT phase and; (2) The magnetic
moment associated with the AF2 structure is independent of pressure
up to the O-cT transition.

Although the neutron powder diffraction measurements demonstrated
the loss of magnetic order in the cT phase, subsequent
first-principles calculations\cite{yildirim} have claimed that the
cT phase is magnetic and  Fe moments order in a AF1 N\'eel state
(all nearest neighbor interactions are AF) with a moment of
1.3~$\mu_\textup{B}$.  There are two three-dimensional realizations
of the AF1 structure, illustrated in Fig.~\ref{fig:fig2}, with
either an antiferromagnetic (Fig.~\ref{fig:fig2}(b)) or
ferromagnetic (Fig.~\ref{fig:fig2}(c)) alignment of adjacent Fe
planes along the c-axis. For both cases, the magnetic unit cell is
the same as the chemical unit cell.  For the structure in
Fig.~\ref{fig:fig2}(b), magnetic reflections will be found at
positions (\textit{h0l})$_{\textup{cT}}$ with \emph{h} and \emph{l}
odd. Unfortunately, these positions also correspond to allowed
nuclear reflections and, unless the moment is large, the magnetic
contribution to the scattering is difficult to measure using
unpolarized neutrons.  For the structure illustrated in
Fig.~\ref{fig:fig2}(c), magnetic reflections will be found at
positions (\textit{h0l})$_{\textup{cT}}$ with \emph{h} odd and
\emph{l} even.  These positions are forbidden for nuclear scattering
from the I4/mmm tetragonal structure and were investigated using the
70 mg FeAs-flux grown sample. Measurements were done at ambient
pressure and T~=~167~K (in the O phase for reference) and at
p~=~0.62~GPa and T~=~50~K (in the cT phase).  We found no evidence
of AF1 magnetic order at the (100)$_{\textup{cT}}$ and
(102)$_{\textup{cT}}$ positions, consistent with our previous
neutron diffraction measurements\cite{kreyssig}.

\subsection{Coexistence of phases in \ca under pressure}
\begin{figure}
\begin{center}
\includegraphics[clip, width=.45\textwidth]{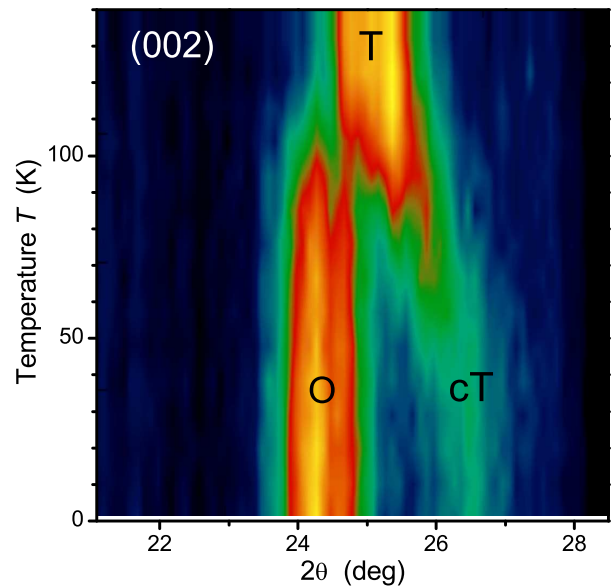}\\
\caption{ (color online) Measurement of the (002) nuclear reflection
from CaFe$_{2}$As$_{2}$ with an area detector on the E4
diffractometer using a Be-Cu clamp-type cell. For an initial
pressure of 0.83 GPa, at room temperature, the T phase transforms to
a mixture of the cT and O phase below approximately 100 K.}
\label{fig:fig4}
\end{center}
\end{figure}

While our single crystal data (Section IIA) clearly shows the
presence of single phase fields in the p-T phase diagram, we do find
circumstances in which extended ranges of coexistence between the cT
and O or T phases occur. We believe that this observation is key to
understanding recent $\mu$SR measurements that suggest that a
partial volume fraction of static magnetic order coexists with
superconductivity over an intermediate pressure range\cite{goko}.
For example, in Fig.~\ref{fig:fig4} we show the evolution of the
(002)$_{\textup{T}}$ diffraction peak as a function of temperature
at 0.83 GPa (set at room temperature) measured on the E4
diffractometer using a Be-Cu clamp-type pressure cell and a 1:1
mixture of Fluorinert 77 and 70, for the pressure medium. On
decreasing the temperature below approximately 120 K (P =
0.66(5)~GPa), we find evidence of an extended coexistence regime
between the T and O phases.  As temperature is further decreased
below approximately 100 K (P = 0.60(5)~GPa), Fig.~\ref{fig:fig4}
shows coexistence between the O and cT structures that persists down
to the base temperature of 2~K.

The coexistence of the antiferromagnetic O and the nonmagnetic cT
phase under conditions similar to those for the $\mu$SR
measurements\cite{goko}, provides a compelling explanation for the
observed coexistence between static magnetic order, with a volume
fraction that decreases with increasing pressure, and a non-magnetic
volume fraction, that increases with increasing pressure. We
attribute this behavior to the freezing of the pressure mediating
liquid above 150K (Daphne oil 7373\cite{murata} in ref
~\onlinecite{goko} and Fluorinert here). This freezing, coupled with
the anisotropy in the change across the T-cT transition, can lead to
significant pressure gradients through the sample(s), a problem that
is not encountered in our He gas cell apparatus for the temperature
and pressure ranges investigated here. As noted in Section IIB,
bonding the sample to a support can result in similar strains and
inhomogeneities, smearing the transitions to and from the cT phase
and extending the coexistence regime between phases.

\section{Discussion}
Although it appears, so far, that the related compounds,
BaFe$_{2}$As$_{2}$ and SrFe$_{2}$As$_{2}$ do not manifest a cT phase
at elevated pressures\cite{kimber}, the transition to a cT phase in
\ca with applied pressure is not unique among systems that
crystallize in the ThCr$_{2}$Si$_{2}$ structure.  Other examples of
this phenomenon are found among the phosphide compounds including
SrRh$_{2}$P$_{2}$ and EuRh$_{2}$P$_{2}$\cite{huhnt1}, as well as
SrNi$_{2}$P$_{2}$, EuCo$_{2}$P$_{2}$\cite{huhnt2} and
EuFe$_{2}$P$_{2}$\cite{bni}.  For the Eu-compounds, the cT phase is
accompanied by a valence transition from Eu$^{2+}$ to the
non-magnetic Eu$^{3+}$ and, for EuCo$_{2}$P$_{2}$, a change from
local moment Eu(4f) to itinerant Co(3d) magnetism associated with a
strong modification of the 3d bands in the cT phase\cite{chefki}. In
all of these phosphide compounds, the striking decrease in the
\textit{c/a} ratios in the cT phase has been described in terms of
"bonding" transitions involving the formation of a P-P single bond
between ions in neighboring planes along the c-axis, and has been
discussed in some detail, by Hoffmann and Zheng\cite{hoffmann}.

Following this work on the small A-site cation limit for the
isostructural phosphide compounds\cite{hoffmann}, we suggest that
there is a transition in the bonding character of As-ions and the
promotion of As-As bonds across the Fe$_{2}$As$_{2}$ layers under
pressure.  This was also suggested in recent theoretical work by
Yildirim\cite{yildirim}. In the ambient pressure tetragonal phase,
the As-As distance between neighboring Fe$_{2}$As$_{2}$ layers is
approximately 3.15~{\AA}, much larger than the As-As single bond
distance of 2.52~{\AA} in elemental As.  In the cT phase, the As-As
distance decreases to approximately 2.82~{\AA}, still larger, but
much closer to the range of a As-As single bond. The enhancement of
the As-As bonding under pressure can have dramatic effects on band
structure and magnetism\cite{hoffmann}. In the case of the collapsed
phase of EuCo$_{2}$P$_{2}$, for example, an increase in the Co 3d
band-filling results in an enhancement of the density of states at
the Fermi energy and the formation of a Co magnetic
moment\cite{bni}.  It has also been pointed out that even small
changes in the arsenic position (z$_{As}$) strongly affects the
occupation of the Fe 3d$_{x^{2}-y^{2}}$ orbitals and, therefore, the
magnetic behavior\cite{krellner}.

To further investigate the impact of the cT transition on the
electronic density of states (DOS) and generalized susceptibility,
$\chi$(q), of \ca, we have performed band structure calculations for
both the T and cT phases. These calculations were performed using
the full potential LAPW method, with
R$_{MT}$~*~K$_{\textrm{max}}$~=~8 , R$_{MT}$~=~2.2, 2.0, 2.0 atomic
units for Ca, Fe and As respectively. The number of k-points in the
irreducible Brillouin zone are 550 for the self consistent charge,
828 for the DOS calculation, and 34501 for the $\chi$(q)
calculation. For the local density functional, the Perdew-Wang 1992
functional\cite{perdew} was employed.  The convergence criterion for
the total energy was 0.01 mRyd/cell. The structural parameters for
the T and cT phases were obtained from experiment\cite{kreyssig}.
The density of states, obtained with the tetrahedron method, has
been broadened with a Gaussian of width 3~mRyd.

\begin{figure}
\begin{center}
\includegraphics[clip, width=.45\textwidth]{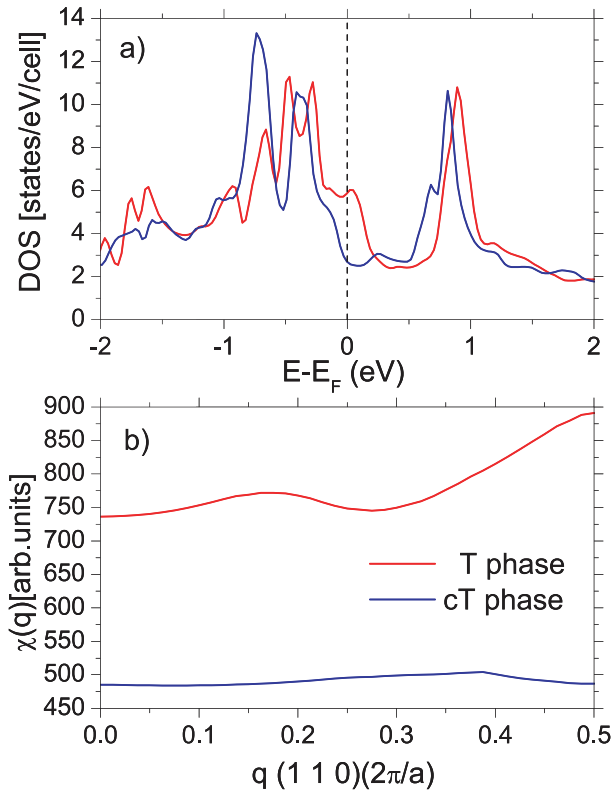}
\caption{(color online) (a) Calculated DOS near the Fermi energy for
the T (red line) and cT (blue line) phases of CaFe$_{2}$As$_{2}$ (b)
The generalized susceptibility $\chi$(q) for the T (red line) and cT
(blue line) phases of CaFe$_{2}$As$_{2}$} \label{fig:fig5}
\end{center}
\end{figure}

Fig.~\ref{fig:fig5}(a) shows the calculated density of states (DOS)
within 2 eV of the Fermi energy for both non-magnetic tetragonal
phases.  The Fe states overwhelmingly ($>$~95 percent) contribute in
this region, with significant As hybridization (bonding) occurring
at lower energies and Ca contributions appearing at higher energies.
The most significant change in the collapsed phase is the dramatic
lowering of the DOS at the Fermi energy associated, primarily, with
a shift to lower energies of the Fe $3d_{x^{2}-y^{2}}$ and Fe
$3d_{xz+yz}$ orbitals. This is also demonstrated in the
corresponding $\chi$(q) calculations shown in Fig.~\ref{fig:fig5}(b)
for the two phases.

The intraband contribution to $\chi$(q~=~0) is exactly the density
of states at $E_{F}$, with a large value favoring ferromagnetic
ordering (Stoner criteria).  We note that the DOS at $E_{F}$ of 2.8
states/ev-Fe is comparable to that for pure Fe (3.0 states/eV-Fe).
The considerably larger peak in $\chi$(q) at the zone boundary for
the T phase, however, is an indication that the magnetic instability
is antiferromagnetic, with ordering observed upon a small
orthorhombic distortion\cite{goldman}. The small DOS at $E_{F}$ for
the cT phase shown in Fig.~\ref{fig:fig5}(a) is not sufficient to
induce magnetic ordering, and the essentially featureless $\chi$(q)
(note the offset) in Fig.~\ref{fig:fig5}(b) and along other high
symmetry directions (not shown) for the cT phase indicates that
magnetic ordering is unlikely at any q-vector.

These calculations clearly support the notion of a depressed Fe 
$3d$ density of states at $E_{F}$, which is consistent with 
previous work\cite{yildirim,krellner,samolyuk} as well as the 
suppression of magnetism in the cT phase. The suppression of the 
DOS realized in these calculations is, however, surprising in 
light of the strong reduction in the resistivity of \ca found 
upon transformations into the cT phase\cite{torikachvili,park}. 
This suggests that scattering effects are greatly reduced in the 
cT phase in comparison to the T phase, and calls for further 
investigation of the spin-fluctuation spectrum of the T phase in 
particular.

\section{Summary and Conclusions}
To summarize our results, we have identified the phase lines
corresponding to transitions between the ambient-pressure
tetragonal, the antiferromagnetic orthorhombic, and the non-magnetic
collapsed tetragonal phases of CaFe$_{2}$As$_{2}$. No additional
structures for pressures up to 2.5 GPa (at 300~K) were observed, in
contradiction to the proposal of Ref~~\onlinecite{lee}. For at least
one of the two possible AF1 structures, we find no evidence of
magnetic ordering in the cT phase as proposed in
Ref.~~\onlinecite{yildirim}.

Whereas the low-pressure T-O transition presents only slight
hysteresis ($\sim$2~K), both the T-cT and O-cT transitions exhibit
very significant hysteresis effects.  The large temperature/pressure
range of hysteresis, together with the strongly anisotropic changes
in the CaFe$_{2}$As$_{2}$ lattice at the cT transition must be
considered in the interpretation of resistivity and susceptibility
measurements in \ca, particularly for measurements where
non-hydrostatic pressure effects are possible.  We have demonstrated
that the $\mu$SR measurements in Ref~~\onlinecite{goko} are
consistent with our neutron powder and single-crystal studies; the
coexistence of non-magnetic and magnetically ordered fractions
results from coexistence between the O and cT phases in the presence
of a non-hydrostatic pressure component. We also note that
pressure-induced superconductivity in CaFe$_{2}$As$_{2}$ has been
observed close to the O-cT transition, and within the hysteretic
region associated with that transition and, therefore, should be
studied carefully under hydrostatic conditions.

The loss of magnetism in the cT phase is largely due to changes in
the band structure that depletes Fe 3$d$ DOS at E$_{F}$, a 
behavior that is analogous to other 122 type 
compounds\cite{huhnt1,huhnt2,bni,hoffmann} and qualitatively 
consistent with previous theoretical 
calculations\cite{yildirim,krellner,samolyuk}.

The authors wish to acknowledge very useful discussions with Joerg
Schmalian, the assistance of J.Q. Yan with sample preparation, and
the assistance of Yejun Feng and Doug Robinson with the high energy
x-ray measurements. The work at the Ames Laboratory and at the MUCAT
sector was supported by the U.S. DOE under Contract No.
DE-AC02-07CH11358. The use of the Advanced Photon Source was
supported by U.S. DOE under Contract No. DE-AC02-06CH113.
\bibliographystyle{apsrev}
\bibliography{cafe2as2}

\begin{thebibliography}{26}
\expandafter\ifx\csname natexlab\endcsname\relax\def\natexlab#1{#1}\fi
\expandafter\ifx\csname bibnamefont\endcsname\relax
  \def\bibnamefont#1{#1}\fi
\expandafter\ifx\csname bibfnamefont\endcsname\relax
  \def\bibfnamefont#1{#1}\fi
\expandafter\ifx\csname citenamefont\endcsname\relax
  \def\citenamefont#1{#1}\fi
\expandafter\ifx\csname url\endcsname\relax
  \def\url#1{\texttt{#1}}\fi
\expandafter\ifx\csname urlprefix\endcsname\relax\def\urlprefix{URL }\fi
\providecommand{\bibinfo}[2]{#2}
\providecommand{\eprint}[2][]{\url{#2}}

\bibitem[{\citenamefont{Torikachvili et~al.}(2008)\citenamefont{Torikachvili,
  Bud'ko, Ni, and Canfield}}]{torikachvili}
\bibinfo{author}{\bibfnamefont{M.~S.} \bibnamefont{Torikachvili}},
  \bibinfo{author}{\bibfnamefont{S.~L.} \bibnamefont{Bud'ko}},
  \bibinfo{author}{\bibfnamefont{N.}~\bibnamefont{Ni}}, \bibnamefont{and}
  \bibinfo{author}{\bibfnamefont{P.~C.} \bibnamefont{Canfield}},
  \bibinfo{journal}{Phys. Rev. Lett.} \textbf{\bibinfo{volume}{101}},
  \bibinfo{pages}{057006} (\bibinfo{year}{2008}).

\bibitem[{\citenamefont{Park et~al.}(2008)\citenamefont{Park, Park, Lee,
  Klimczuk, Bauer, Ronning, and Thompson}}]{park}
\bibinfo{author}{\bibfnamefont{T.}~\bibnamefont{Park}},
  \bibinfo{author}{\bibfnamefont{E.}~\bibnamefont{Park}},
  \bibinfo{author}{\bibfnamefont{H.}~\bibnamefont{Lee}},
  \bibinfo{author}{\bibfnamefont{T.}~\bibnamefont{Klimczuk}},
  \bibinfo{author}{\bibfnamefont{E.~D.} \bibnamefont{Bauer}},
  \bibinfo{author}{\bibfnamefont{F.}~\bibnamefont{Ronning}}, \bibnamefont{and}
  \bibinfo{author}{\bibfnamefont{J.~D.} \bibnamefont{Thompson}},
  \bibinfo{journal}{J. Phys.: Condens. Matter} \textbf{\bibinfo{volume}{20}},
  \bibinfo{pages}{322204} (\bibinfo{year}{2008}).

\bibitem[{\citenamefont{Norman}(2008)}]{norman}
\bibinfo{author}{\bibfnamefont{M.~R.} \bibnamefont{Norman}},
  \bibinfo{journal}{Physics} \textbf{\bibinfo{volume}{1}}, \bibinfo{pages}{21}
  (\bibinfo{year}{2008}), \bibinfo{note}{and references therein}.

\bibitem[{\citenamefont{Okada et~al.}(2008)\citenamefont{Okada, Igawa,
  Takahashi, Kamihara, Hirano, Hosono, Matsubayashi, and Uwatoko}}]{okada}
\bibinfo{author}{\bibfnamefont{H.}~\bibnamefont{Okada}},
  \bibinfo{author}{\bibfnamefont{K.}~\bibnamefont{Igawa}},
  \bibinfo{author}{\bibfnamefont{H.}~\bibnamefont{Takahashi}},
  \bibinfo{author}{\bibfnamefont{Y.}~\bibnamefont{Kamihara}},
  \bibinfo{author}{\bibfnamefont{M.}~\bibnamefont{Hirano}},
  \bibinfo{author}{\bibfnamefont{H.}~\bibnamefont{Hosono}},
  \bibinfo{author}{\bibfnamefont{K.}~\bibnamefont{Matsubayashi}},
  \bibnamefont{and} \bibinfo{author}{\bibfnamefont{Y.}~\bibnamefont{Uwatoko}},
  \bibinfo{journal}{J. Phys. Soc. Jpn.} \textbf{\bibinfo{volume}{77}},
  \bibinfo{pages}{113712} (\bibinfo{year}{2008}).

\bibitem[{\citenamefont{Huang et~al.}(2008)\citenamefont{Huang, Qiu, Bao,
  Green, Lynn, Gasparovic, Wu, Wu, and Chen}}]{huang}
\bibinfo{author}{\bibfnamefont{Q.}~\bibnamefont{Huang}},
  \bibinfo{author}{\bibfnamefont{Y.}~\bibnamefont{Qiu}},
  \bibinfo{author}{\bibfnamefont{W.}~\bibnamefont{Bao}},
  \bibinfo{author}{\bibfnamefont{M.~A.} \bibnamefont{Green}},
  \bibinfo{author}{\bibfnamefont{J.~W.} \bibnamefont{Lynn}},
  \bibinfo{author}{\bibfnamefont{Y.~C.} \bibnamefont{Gasparovic}},
  \bibinfo{author}{\bibfnamefont{T.}~\bibnamefont{Wu}},
  \bibinfo{author}{\bibfnamefont{G.}~\bibnamefont{Wu}}, \bibnamefont{and}
  \bibinfo{author}{\bibfnamefont{X.~H.} \bibnamefont{Chen}},
  \bibinfo{journal}{arXiv:0806.2776}  (\bibinfo{year}{2008}),
  \bibinfo{note}{unpublished}.

\bibitem[{\citenamefont{Jesche et~al.}(2008)\citenamefont{Jesche,
  Caroca-Canales, Rosner, Borrmann, Ormeci, Kasinathan, Kaneko, Klauss,
  Luetkens, Khasanov et~al.}}]{jesche}
\bibinfo{author}{\bibfnamefont{A.}~\bibnamefont{Jesche}},
  \bibinfo{author}{\bibfnamefont{N.}~\bibnamefont{Caroca-Canales}},
  \bibinfo{author}{\bibfnamefont{H.}~\bibnamefont{Rosner}},
  \bibinfo{author}{\bibfnamefont{H.}~\bibnamefont{Borrmann}},
  \bibinfo{author}{\bibfnamefont{A.}~\bibnamefont{Ormeci}},
  \bibinfo{author}{\bibfnamefont{D.}~\bibnamefont{Kasinathan}},
  \bibinfo{author}{\bibfnamefont{K.}~\bibnamefont{Kaneko}},
  \bibinfo{author}{\bibfnamefont{H.~H.} \bibnamefont{Klauss}},
  \bibinfo{author}{\bibfnamefont{H.}~\bibnamefont{Luetkens}},
  \bibinfo{author}{\bibfnamefont{R.}~\bibnamefont{Khasanov}},
  \bibnamefont{et~al.}, \bibinfo{journal}{arXiv:0807.0632}
  (\bibinfo{year}{2008}), \bibinfo{note}{unpublished}.

\bibitem[{\citenamefont{Zhao et~al.}(2008)\citenamefont{Zhao, Ratcliff, II,
  Lynn, Chen, Luo, Wang, Hu, and Dai}}]{zhao}
\bibinfo{author}{\bibfnamefont{J.}~\bibnamefont{Zhao}},
  \bibinfo{author}{\bibfnamefont{W.}~\bibnamefont{Ratcliff}},
  \bibinfo{author}{\bibnamefont{II}}, \bibinfo{author}{\bibfnamefont{J.~W.}
  \bibnamefont{Lynn}}, \bibinfo{author}{\bibfnamefont{G.~F.}
  \bibnamefont{Chen}}, \bibinfo{author}{\bibfnamefont{J.~L.}
  \bibnamefont{Luo}}, \bibinfo{author}{\bibfnamefont{N.~L.}
  \bibnamefont{Wang}}, \bibinfo{author}{\bibfnamefont{J.}~\bibnamefont{Hu}},
  \bibnamefont{and} \bibinfo{author}{\bibfnamefont{P.}~\bibnamefont{Dai}},
  \bibinfo{journal}{Phys. Rev. B} \textbf{\bibinfo{volume}{78}},
  \bibinfo{pages}{140504(R)} (\bibinfo{year}{2008}).

\bibitem[{\citenamefont{Su et~al.}(2008)\citenamefont{Su, Link, Schneidewind,
  Wolf, Xiao, Mittal, Rotter, Johrendt, Brueckel, and Loewenhaupt}}]{su}
\bibinfo{author}{\bibfnamefont{Y.}~\bibnamefont{Su}},
  \bibinfo{author}{\bibfnamefont{P.}~\bibnamefont{Link}},
  \bibinfo{author}{\bibfnamefont{A.}~\bibnamefont{Schneidewind}},
  \bibinfo{author}{\bibfnamefont{T.}~\bibnamefont{Wolf}},
  \bibinfo{author}{\bibfnamefont{Y.}~\bibnamefont{Xiao}},
  \bibinfo{author}{\bibfnamefont{R.}~\bibnamefont{Mittal}},
  \bibinfo{author}{\bibfnamefont{M.}~\bibnamefont{Rotter}},
  \bibinfo{author}{\bibfnamefont{D.}~\bibnamefont{Johrendt}},
  \bibinfo{author}{\bibfnamefont{T.}~\bibnamefont{Brueckel}}, \bibnamefont{and}
  \bibinfo{author}{\bibfnamefont{M.}~\bibnamefont{Loewenhaupt}},
  \bibinfo{journal}{arXiv:0807.1743}  (\bibinfo{year}{2008}),
  \bibinfo{note}{unpublished}.

\bibitem[{\citenamefont{Ni et~al.}(2008)\citenamefont{Ni, Nandi, Kreyssig,
  Goldman, Mun, Bud'ko, and Canfield}}]{ni}
\bibinfo{author}{\bibfnamefont{N.}~\bibnamefont{Ni}},
  \bibinfo{author}{\bibfnamefont{S.}~\bibnamefont{Nandi}},
  \bibinfo{author}{\bibfnamefont{A.}~\bibnamefont{Kreyssig}},
  \bibinfo{author}{\bibfnamefont{A.~I.} \bibnamefont{Goldman}},
  \bibinfo{author}{\bibfnamefont{E.~D.} \bibnamefont{Mun}},
  \bibinfo{author}{\bibfnamefont{S.~L.} \bibnamefont{Bud'ko}},
  \bibnamefont{and} \bibinfo{author}{\bibfnamefont{P.~C.}
  \bibnamefont{Canfield}}, \bibinfo{journal}{Phys. Rev. B}
  \textbf{\bibinfo{volume}{78}}, \bibinfo{pages}{014523}
  (\bibinfo{year}{2008}).

\bibitem[{\citenamefont{Goldman et~al.}(2008)\citenamefont{Goldman, Argyriou,
  Ouladdiaf, Chatterji, Kreyssig, Nandi, Ni, Bud'ko, Canfield, and
  McQueeney}}]{goldman}
\bibinfo{author}{\bibfnamefont{A.~I.} \bibnamefont{Goldman}},
  \bibinfo{author}{\bibfnamefont{D.~N.} \bibnamefont{Argyriou}},
  \bibinfo{author}{\bibfnamefont{B.}~\bibnamefont{Ouladdiaf}},
  \bibinfo{author}{\bibfnamefont{T.}~\bibnamefont{Chatterji}},
  \bibinfo{author}{\bibfnamefont{A.}~\bibnamefont{Kreyssig}},
  \bibinfo{author}{\bibfnamefont{S.}~\bibnamefont{Nandi}},
  \bibinfo{author}{\bibfnamefont{N.}~\bibnamefont{Ni}},
  \bibinfo{author}{\bibfnamefont{S.~L.} \bibnamefont{Bud'ko}},
  \bibinfo{author}{\bibfnamefont{P.~C.} \bibnamefont{Canfield}},
  \bibnamefont{and} \bibinfo{author}{\bibfnamefont{R.~J.}
  \bibnamefont{McQueeney}}, \bibinfo{journal}{Phys. Rev. B}
  \textbf{\bibinfo{volume}{78}}, \bibinfo{pages}{100506(R)}
  (\bibinfo{year}{2008}).

\bibitem[{\citenamefont{Yildirim}(2008{\natexlab{a}})}]{yildirim2}
\bibinfo{author}{\bibfnamefont{T.}~\bibnamefont{Yildirim}},
  \bibinfo{journal}{Phys. Rev. Lett.} \textbf{\bibinfo{volume}{101}},
  \bibinfo{pages}{057010} (\bibinfo{year}{2008}{\natexlab{a}}).

\bibitem[{\citenamefont{Kreyssig et~al.}(2008)\citenamefont{Kreyssig, Green,
  Lee, Samolyuk, Zajdel, Lynn, Bud'ko, Torikachvili, Ni, Nandi
  et~al.}}]{kreyssig}
\bibinfo{author}{\bibfnamefont{A.}~\bibnamefont{Kreyssig}},
  \bibinfo{author}{\bibfnamefont{M.~A.} \bibnamefont{Green}},
  \bibinfo{author}{\bibfnamefont{Y.}~\bibnamefont{Lee}},
  \bibinfo{author}{\bibfnamefont{G.~D.} \bibnamefont{Samolyuk}},
  \bibinfo{author}{\bibfnamefont{P.}~\bibnamefont{Zajdel}},
  \bibinfo{author}{\bibfnamefont{J.~W.} \bibnamefont{Lynn}},
  \bibinfo{author}{\bibfnamefont{S.~L.} \bibnamefont{Bud'ko}},
  \bibinfo{author}{\bibfnamefont{M.~S.} \bibnamefont{Torikachvili}},
  \bibinfo{author}{\bibfnamefont{N.}~\bibnamefont{Ni}},
  \bibinfo{author}{\bibfnamefont{S.}~\bibnamefont{Nandi}},
  \bibnamefont{et~al.}, \bibinfo{journal}{arXiv:0807.3032}
  (\bibinfo{year}{2008}), \bibinfo{note}{unpublished}.

\bibitem[{\citenamefont{Lee et~al.}(2008)\citenamefont{Lee, Park, Park,
  Ronning, Bauer, and Thompson}}]{lee}
\bibinfo{author}{\bibfnamefont{H.}~\bibnamefont{Lee}},
  \bibinfo{author}{\bibfnamefont{E.}~\bibnamefont{Park}},
  \bibinfo{author}{\bibfnamefont{T.}~\bibnamefont{Park}},
  \bibinfo{author}{\bibfnamefont{F.}~\bibnamefont{Ronning}},
  \bibinfo{author}{\bibfnamefont{E.~D.} \bibnamefont{Bauer}}, \bibnamefont{and}
  \bibinfo{author}{\bibfnamefont{J.~D.} \bibnamefont{Thompson}},
  \bibinfo{journal}{arXiv:0809.3550}  (\bibinfo{year}{2008}),
  \bibinfo{note}{unpublished}.

\bibitem[{\citenamefont{Goko et~al.}(2008)\citenamefont{Goko, Aczel,
  Baggio-Saitovitch, Bud'ko, Canfield, Carlo, Chen, Dai, Hamann, Hu
  et~al.}}]{goko}
\bibinfo{author}{\bibfnamefont{T.}~\bibnamefont{Goko}},
  \bibinfo{author}{\bibfnamefont{A.~A.} \bibnamefont{Aczel}},
  \bibinfo{author}{\bibfnamefont{E.}~\bibnamefont{Baggio-Saitovitch}},
  \bibinfo{author}{\bibfnamefont{S.~L.} \bibnamefont{Bud'ko}},
  \bibinfo{author}{\bibfnamefont{P.}~\bibnamefont{Canfield}},
  \bibinfo{author}{\bibfnamefont{J.~P.} \bibnamefont{Carlo}},
  \bibinfo{author}{\bibfnamefont{G.~F.} \bibnamefont{Chen}},
  \bibinfo{author}{\bibfnamefont{P.}~\bibnamefont{Dai}},
  \bibinfo{author}{\bibfnamefont{A.~C.} \bibnamefont{Hamann}},
  \bibinfo{author}{\bibfnamefont{W.~Z.} \bibnamefont{Hu}},
  \bibnamefont{et~al.}, \bibinfo{journal}{arXiv:0808.1425}
  (\bibinfo{year}{2008}), \bibinfo{note}{unpublished}.

\bibitem[{\citenamefont{Yildirim}(2008{\natexlab{b}})}]{yildirim}
\bibinfo{author}{\bibfnamefont{T.}~\bibnamefont{Yildirim}},
  \bibinfo{journal}{arXiv:0807.3936}  (\bibinfo{year}{2008}{\natexlab{b}}),
  \bibinfo{note}{unpublished}.

\bibitem[{\citenamefont{Kreyssig et~al.}(2007)\citenamefont{Kreyssig, Chang,
  Janssen, Kim, Nandi, Yan, Tan, McQueeney, Canfield, and Goldman}}]{kreyssig2}
\bibinfo{author}{\bibfnamefont{A.}~\bibnamefont{Kreyssig}},
  \bibinfo{author}{\bibfnamefont{S.}~\bibnamefont{Chang}},
  \bibinfo{author}{\bibfnamefont{Y.}~\bibnamefont{Janssen}},
  \bibinfo{author}{\bibfnamefont{J.~W.} \bibnamefont{Kim}},
  \bibinfo{author}{\bibfnamefont{S.}~\bibnamefont{Nandi}},
  \bibinfo{author}{\bibfnamefont{J.~Q.} \bibnamefont{Yan}},
  \bibinfo{author}{\bibfnamefont{L.}~\bibnamefont{Tan}},
  \bibinfo{author}{\bibfnamefont{R.~J.} \bibnamefont{McQueeney}},
  \bibinfo{author}{\bibfnamefont{P.~C.} \bibnamefont{Canfield}},
  \bibnamefont{and} \bibinfo{author}{\bibfnamefont{A.~I.}
  \bibnamefont{Goldman}}, \bibinfo{journal}{Phys. Rev. B}
  \textbf{\bibinfo{volume}{76}}, \bibinfo{pages}{054421}
  (\bibinfo{year}{2007}).

\bibitem[{\citenamefont{Murata et~al.}(1997)\citenamefont{Murata, Yoshino,
  Yadav, Honda, and Shirakawa}}]{murata}
\bibinfo{author}{\bibfnamefont{K.}~\bibnamefont{Murata}},
  \bibinfo{author}{\bibfnamefont{H.}~\bibnamefont{Yoshino}},
  \bibinfo{author}{\bibfnamefont{H.~O.} \bibnamefont{Yadav}},
  \bibinfo{author}{\bibfnamefont{Y.}~\bibnamefont{Honda}}, \bibnamefont{and}
  \bibinfo{author}{\bibfnamefont{N.}~\bibnamefont{Shirakawa}},
  \bibinfo{journal}{Rev. Sci. Instrum.} \textbf{\bibinfo{volume}{68}},
  \bibinfo{pages}{2490} (\bibinfo{year}{1997}).

\bibitem[{\citenamefont{Kimber et~al.}(2008)\citenamefont{Kimber, Kreyssig,
  Yokaichiya, Argyriou, Yan, Hansen, Chatterji, McQueeney, Canfield, and
  Goldman}}]{kimber}
\bibinfo{author}{\bibfnamefont{S.~A.~J.} \bibnamefont{Kimber}},
  \bibinfo{author}{\bibfnamefont{A.}~\bibnamefont{Kreyssig}},
  \bibinfo{author}{\bibfnamefont{F.}~\bibnamefont{Yokaichiya}},
  \bibinfo{author}{\bibfnamefont{D.~N.} \bibnamefont{Argyriou}},
  \bibinfo{author}{\bibfnamefont{J.~Q.} \bibnamefont{Yan}},
  \bibinfo{author}{\bibfnamefont{T.}~\bibnamefont{Hansen}},
  \bibinfo{author}{\bibfnamefont{T.}~\bibnamefont{Chatterji}},
  \bibinfo{author}{\bibfnamefont{R.~J.} \bibnamefont{McQueeney}},
  \bibinfo{author}{\bibfnamefont{P.~C.} \bibnamefont{Canfield}},
  \bibnamefont{and} \bibinfo{author}{\bibfnamefont{A.~I.}
  \bibnamefont{Goldman}} (\bibinfo{year}{2008}), \bibinfo{note}{unpublished}.

\bibitem[{\citenamefont{Huhnt et~al.}(1997{\natexlab{a}})\citenamefont{Huhnt,
  Michels, Roepke, Schlabitz, Wurth, Johrendt, and Mewis}}]{huhnt1}
\bibinfo{author}{\bibfnamefont{C.}~\bibnamefont{Huhnt}},
  \bibinfo{author}{\bibfnamefont{G.}~\bibnamefont{Michels}},
  \bibinfo{author}{\bibfnamefont{M.}~\bibnamefont{Roepke}},
  \bibinfo{author}{\bibfnamefont{W.}~\bibnamefont{Schlabitz}},
  \bibinfo{author}{\bibfnamefont{A.}~\bibnamefont{Wurth}},
  \bibinfo{author}{\bibfnamefont{D.}~\bibnamefont{Johrendt}}, \bibnamefont{and}
  \bibinfo{author}{\bibfnamefont{A.}~\bibnamefont{Mewis}},
  \bibinfo{journal}{Physica B} \textbf{\bibinfo{volume}{240}},
  \bibinfo{pages}{26} (\bibinfo{year}{1997}{\natexlab{a}}).

\bibitem[{\citenamefont{Huhnt et~al.}(1997{\natexlab{b}})\citenamefont{Huhnt,
  Schlabitz, Wurth, Mewis, and Reehuis}}]{huhnt2}
\bibinfo{author}{\bibfnamefont{C.}~\bibnamefont{Huhnt}},
  \bibinfo{author}{\bibfnamefont{W.}~\bibnamefont{Schlabitz}},
  \bibinfo{author}{\bibfnamefont{A.}~\bibnamefont{Wurth}},
  \bibinfo{author}{\bibfnamefont{A.}~\bibnamefont{Mewis}}, \bibnamefont{and}
  \bibinfo{author}{\bibfnamefont{M.}~\bibnamefont{Reehuis}},
  \bibinfo{journal}{Phys. Rev. B} \textbf{\bibinfo{volume}{56}},
  \bibinfo{pages}{13796} (\bibinfo{year}{1997}{\natexlab{b}}).

\bibitem[{\citenamefont{Ni et~al.}(2001)\citenamefont{Ni, Abd-Elmeguid,
  Micklitz, Sanchez, Vulliet, and Johrendt}}]{bni}
\bibinfo{author}{\bibfnamefont{B.}~\bibnamefont{Ni}},
  \bibinfo{author}{\bibfnamefont{M.~M.} \bibnamefont{Abd-Elmeguid}},
  \bibinfo{author}{\bibfnamefont{H.}~\bibnamefont{Micklitz}},
  \bibinfo{author}{\bibfnamefont{J.~P.} \bibnamefont{Sanchez}},
  \bibinfo{author}{\bibfnamefont{P.}~\bibnamefont{Vulliet}}, \bibnamefont{and}
  \bibinfo{author}{\bibfnamefont{D.}~\bibnamefont{Johrendt}},
  \bibinfo{journal}{Phys. Rev. B} \textbf{\bibinfo{volume}{63}},
  \bibinfo{pages}{100102(R)} (\bibinfo{year}{2001}).

\bibitem[{\citenamefont{Chefki et~al.}(1998)\citenamefont{Chefki, Abd-Elmeguid,
  Micklitz, Huhnt, Schlabitz, Reehuis, and Jeitschko}}]{chefki}
\bibinfo{author}{\bibfnamefont{M.}~\bibnamefont{Chefki}},
  \bibinfo{author}{\bibfnamefont{M.~M.} \bibnamefont{Abd-Elmeguid}},
  \bibinfo{author}{\bibfnamefont{H.}~\bibnamefont{Micklitz}},
  \bibinfo{author}{\bibfnamefont{C.}~\bibnamefont{Huhnt}},
  \bibinfo{author}{\bibfnamefont{W.}~\bibnamefont{Schlabitz}},
  \bibinfo{author}{\bibfnamefont{M.}~\bibnamefont{Reehuis}}, \bibnamefont{and}
  \bibinfo{author}{\bibfnamefont{W.}~\bibnamefont{Jeitschko}},
  \bibinfo{journal}{Phys. Rev. Lett.} \textbf{\bibinfo{volume}{80}},
  \bibinfo{pages}{802} (\bibinfo{year}{1998}).

\bibitem[{\citenamefont{Hoffmann and Zheng}(1985)}]{hoffmann}
\bibinfo{author}{\bibfnamefont{R.}~\bibnamefont{Hoffmann}} \bibnamefont{and}
  \bibinfo{author}{\bibfnamefont{C.}~\bibnamefont{Zheng}}, \bibinfo{journal}{J.
  Phys. Chem.} \textbf{\bibinfo{volume}{89}}, \bibinfo{pages}{4175}
  (\bibinfo{year}{1985}).

\bibitem[{\citenamefont{Krellner et~al.}(2008)\citenamefont{Krellner,
  Caroca-Canales, Jesche, Rosner, Ormeci, and Geibel}}]{krellner}
\bibinfo{author}{\bibfnamefont{C.}~\bibnamefont{Krellner}},
  \bibinfo{author}{\bibfnamefont{N.}~\bibnamefont{Caroca-Canales}},
  \bibinfo{author}{\bibfnamefont{A.}~\bibnamefont{Jesche}},
  \bibinfo{author}{\bibfnamefont{H.}~\bibnamefont{Rosner}},
  \bibinfo{author}{\bibfnamefont{A.}~\bibnamefont{Ormeci}}, \bibnamefont{and}
  \bibinfo{author}{\bibfnamefont{C.}~\bibnamefont{Geibel}},
  \bibinfo{journal}{Phys. Rev. B} \textbf{\bibinfo{volume}{78}},
  \bibinfo{eid}{100504(R)} (\bibinfo{year}{2008}).

\bibitem[{\citenamefont{Perdew and Wang}(1992)}]{perdew}
\bibinfo{author}{\bibfnamefont{J.~P.} \bibnamefont{Perdew}} \bibnamefont{and}
  \bibinfo{author}{\bibfnamefont{Y.}~\bibnamefont{Wang}},
  \bibinfo{journal}{Phys. Rev. B} \textbf{\bibinfo{volume}{45}},
  \bibinfo{pages}{13244} (\bibinfo{year}{1992}).

\bibitem[{\citenamefont{Samolyuk and Antropov}(2008)}]{samolyuk}
\bibinfo{author}{\bibfnamefont{G.~D.} \bibnamefont{Samolyuk}} \bibnamefont{and}
  \bibinfo{author}{\bibfnamefont{V.~P.} \bibnamefont{Antropov}},
  \bibinfo{journal}{arXiv:0810.1445}  (\bibinfo{year}{2008}),
  \bibinfo{note}{unpublished}.

\end{thebibliography}
\end{document}